\documentclass[11pt]{article}
\pdfoutput=1

\usepackage{amsmath}
\usepackage{amssymb}
\usepackage{graphicx}
\usepackage{color}

\usepackage{hyperref}

\def\figdir{figures}

\begin{document}
\title{Indirect detonation initiation using acoustic \\timescale thermal power deposition}
\author{J.~D. Regele\\
{\normalsize\itshape
   Iowa State University, Ames, IA, 50011, USA}\\
D.~R. Kassoy,\ A.~Vezolainen,\ and O.~V. Vasilyev \\
{\normalsize\itshape
   University of Colorado, Boulder, CO 80309, USA}
}

\maketitle
\begin{abstract}
A fluid dynamics video is presented that demonstrates an indirect detonation 
initiation process. In this process, a transient power deposition adds heat 
to a spatially resolved volume of fluid 
in an amount of time that is similar to the acoustic timescale of the fluid volume. 
A highly resolved two-dimensional simulation shows the events that unfold
after the heat is added. 
\end{abstract}

Traditionally, combustion modelers have concluded that detonations form either by
direct initiation or by Deflagration-to-Detonation Transition (DDT).
Direct initiation is accomplished by depositing a large amount 
of energy deposited in a short time period such that a blast wave is created inside 
the reactive gas mixture. In DDT, diffusion, viscosity, 
and turbulence play a major role in preheating the reactive mixture to facilitate
detonation formation
~\cite{Dorofeev2011a,Mazaheri2012,Oran2007,Radulescu2007,Romick2012}. These transport
processes have little or no effect in direct initiation.

Direct initiation and DDT can be seen as two limiting extremes on a continuum scale using the
acoustic timescale theory of Kassoy~\cite{Kassoy2010}. Consider a fluid volume
of length scale $l$ and sound speed $a$ such that the acoustic timescale of the 
fluid volume can be defined $t_a=l/a$. If heat is added to the fluid volume on a 
timescale $t_h$ that is short compared to the acoustic timescale $t_h \ll t_a$
then the fluid experiences nearly constant volume heat addition. The amount of energy added 
to the volume determines whether it will form acoustic, shock, or blast 
waves. 

The numerical simulation presented in this video focuses on acoustic 
timescale detonation initiation, where
 heat is deposited on a timescale similar to the acoustic timescale $t_h \sim t_a$. 
The two-dimensional simulation presented in this video begins with a 
reactive mixture initially at rest and heat is added to a circular volume 
on the acoustic timescale $t_h \sim t_a$.


The simulation begins with the reactive gas at rest with the initial condition
$\rho_0=p_0=Y_0=1$ and $\mathbf{v_0}=0$.
The domain lies in $x \in [-3,57]$ and $y \in [-3,12]$ and reflecting slip walls
are present on all walls except the exit $x=57$.
The heat addition is limited to a circle of radius $R=2$ centered at the origin. 
The simulation uses a heat of reaction $q=15$, specific heat ratio $\gamma=1.4$,
activation energy $E=13.8$, and pre-exponential factor $B=35$.
Heat is added between $t_a=0.5$ and $t_b=5.25$.


The Parallel Adaptive Wavelet-Collocation Method
(PAWCM) is used to capture the wide range of scales that are present
\cite{Kevlahan2005, Vasilyev2000}.
The PAWCM combines second generation wavelets with a prescribed
 error threshold parameter $\epsilon$ to determine which grid points
are necessary in order to achieve a prescribed level of accuracy.
The hyperbolic solver developed for the PAWCM is used to maintain 
numerical stability and reduce spurious oscillations across jump
discontinuities~\cite{Regele2009}. 
The effective grid resolution for the simulation is $15360\times3072$.

\begin{figure}[tb]
 \centering
   \includegraphics[width=0.35\textwidth]{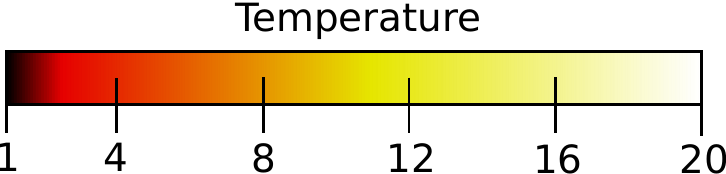}
 \begin{tabular}{cc}
   \includegraphics[width=0.465\textwidth]{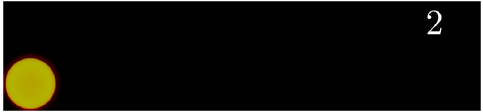} &
   \includegraphics[width=0.465\textwidth]{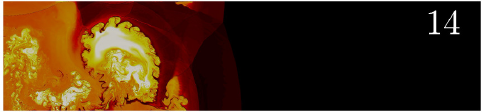} \\
   \includegraphics[width=0.465\textwidth]{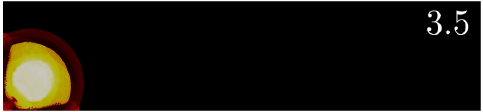} &
   \includegraphics[width=0.465\textwidth]{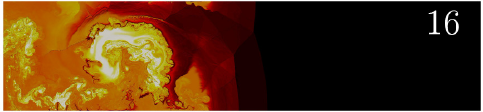} \\
   \includegraphics[width=0.465\textwidth]{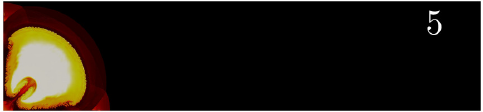} &
   \includegraphics[width=0.465\textwidth]{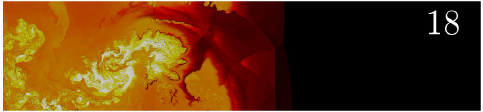} \\
   \includegraphics[width=0.465\textwidth]{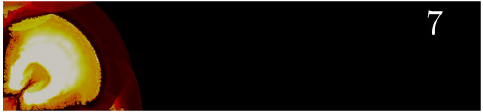} &
   \includegraphics[width=0.465\textwidth]{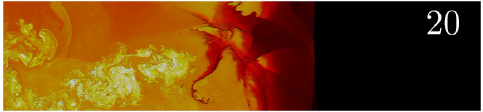} \\
   \includegraphics[width=0.465\textwidth]{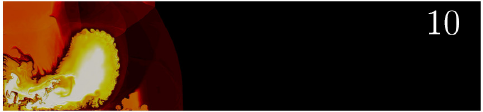} &
   \includegraphics[width=0.465\textwidth]{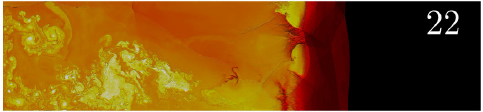} \\
   \includegraphics[width=0.465\textwidth]{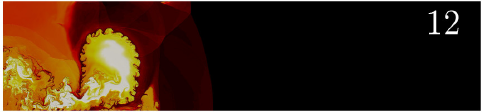} &
   \includegraphics[width=0.465\textwidth]{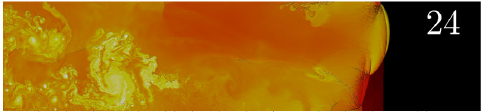} \\
   \includegraphics[width=0.46\textwidth]{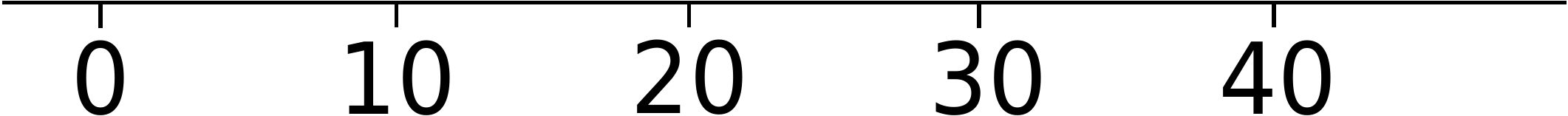} &
  \includegraphics[width=0.46\textwidth]{\figdir/axesbottom2}
 \end{tabular}
\caption{Sequence of temperature contours demonstrate the multidimensional indirect 
detonation formation process for times $2\le t \le 24$.}
\label{fig:2D_j9}
\end{figure}

Figure \ref{fig:2D_j9} demonstrates the indirect detonation initiation
process by presenting a series of snapshots of temperature contours
corresponding to the same events shown in the video. 
Heat is deposited in a circle in the bottom left hand corner
from $0.5\le t \le 5.25$. The rapid deposition of heat creates compression waves
that propagate away from the initially heated region. 
Before $t=2$, the reactants inside the deposition region are
consumed in a chemical explosion, which adds additional heat to the deposition
region. 
It is difficult to discern from the contour at $t=2$, but the interface 
between the burning and reactive gas forms a rippled surface. It is thought that 
this is a result of the Darrieus-Landau instability at the burning gas interface
because once the reactants are consumed the growth of surface fluctuations ceases.
%
%

At some point shortly after $t=2$ when the compression waves first reflect off
the left and bottom walls the compression waves become fully discontinuous shock
waves. The reflected and transmitted shocks form Mach stems that propagate in
the positive $x$- and $y$-directions. The reflected waves impinge on the 
burnt-unburnt gas interface and induce Richtmyer-Meshkov instabilities, which then
increase the fluctuation magnitude at the material interface. This can be initially
observed in the $t=3.5$ contour.

At about $t=2.5$, a second explosion occurs in the lower left corner
when the shock waves reflect off the bottom and left walls and 
raise the pressure in that region in a duration short enough
that the temperature rises with pressure. The reactive gas explodes once it has
reached a sufficient temperature. 

In the $t=5$ frame, the original outward propagating shock wave has reflected off the
left and bottom walls and the Mach stems are clearly visible in the temperature
contour. On the left wall, the leading edge of the shock wave is just about to
reflect off the upper wall, starting in the upper left corner.
When reflection occurs on the upper boundary, a hot spot appears in the upper
left-hand corner of the channel, characterized by substantial local inertial
confinement. This hot spot releases heat and generates compression waves that 
propagate away from the hot spot location. 

At $t=7$, Fig.~\ref{fig:2D_j9} shows the reflected wave re-enters the reacted region and
is refracted, which induces an additional longitudinal component to the wave direction.
The transverse waves compress and heat previously unreacted fuel pockets, which ignite
and help produce additional longitudinal waves, as well as sustaining the transverse 
waves the reverberate off the top and bottom walls.

Kelvin-Helmholtz roll-up instabilities are clearly visible in frames $t=[10, 12, 14]$
at the burnt-unburnt gas interface with a fairly high level of detail. The 
existence of such a detailed interface serves as an indicator that any numerical
diffusion present in the algorithm has been minimized to the point that these 
features are possible to capture.

At about $t=14$ the heat release rate by the preheated gas begins to escalate.
This acceleration in
heat release can be observed in the temperature contour sequence as the rapid 
consumption of fuel starting at $t=14$ and ending at $t=24$ with the formation of 
the over-driven detonation wave emerging from the lead shock front.

\section*{Acknowledgements}
J.D.R. would like to thank Guillaume Blanquart for the use of his computational
resources to perform this simulation.

\bibliographystyle{plain}
\bibliography{library}

\begin{thebibliography}{1}

\bibitem{Dorofeev2011a}
Sergey~B. Dorofeev.
\newblock {Flame acceleration and explosion safety applications}.
\newblock {\em Proceedings of the Combustion Institute}, 33(2):2161--2175,
  January 2011.

\bibitem{Kassoy2010}
D.~R. Kassoy.
\newblock {The response of a compressible gas to extremely rapid transient,
  spatially resolved energy addition: an asymptotic formulation}.
\newblock {\em Journal of Engineering Mathematics}, 68(3-4):249--262, August
  2010.

\bibitem{Kevlahan2005}
N~K~R Kevlahan and O~V Vasilyev.
\newblock {An adaptive wavelet collocation method for fluid-structure
  interaction at high reynolds numbers}.
\newblock {\em Sciam J Sci Comput}, 26(6):1894--1915, 2005.

\bibitem{Mazaheri2012}
Kiumars Mazaheri, Yasser Mahmoudi, and Matei~I. Radulescu.
\newblock {Diffusion and hydrodynamic instabilities in gaseous detonations}.
\newblock {\em Combustion and Flame}, 159(6):2138--2154, June 2012.

\bibitem{Oran2007}
Elaine~S. Oran and Vadim~N. Gamezo.
\newblock {Origins of the deflagration-to-detonation transition in gas-phase
  combustion}.
\newblock {\em Combustion and Flame}, 148(1-2):4--47, January 2007.

\bibitem{Radulescu2007}
Matei~I. Radulescu, Gary~J. Sharpe, Chung~K. Law, and John H.~S. Lee.
\newblock {The hydrodynamic structure of unstable cellular detonations}.
\newblock {\em Journal of Fluid Mechanics}, 580:31, May 2007.

\bibitem{Regele2009}
J~D Regele and O~V Vasilyev.
\newblock {An Adaptive Wavelet-Collocation Method for Shock Computations}.
\newblock {\em International Journal of Computational Fluid Dynamics},
  23(7):503--518, 2009.

\bibitem{Romick2012}
C.~M. Romick, T.~D. Aslam, and J.~M. Powers.
\newblock {The effect of diffusion on the dynamics of unsteady detonations}.
\newblock {\em Journal of Fluid Mechanics}, 699:453--464, April 2012.

\bibitem{Vasilyev2000}
Oleg~V Vasilyev and Christopher Bowman.
\newblock {Second-Generation Wavelet Collocation Method for the Solution of
  Partial Differential Equations}.
\newblock {\em Journal of Computational Physics}, 165(2):660--693, December
  2000.

\end{thebibliography}

\end{document}